\documentclass[11pt,twoside]{book}
\usepackage{vshalo}
\usepackage{longtable}
\usepackage{amsmath,amssymb}
\usepackage{graphicx}
\usepackage{lscape}
\usepackage{psfig}
\usepackage{epsf}
\usepackage{index}
\usepackage{natbib}
\usepackage{bigdelim}
\usepackage{multirow}
\makeindex
\newindex{aut}{adx}{and}{\Large Author  Index}
\newcommand{\axindex}[1]{\index[aut]{#1}}

\begin{document}

\pagestyle{myheadings}
\markboth{Sharina et al.}{Globular clusters in dwarf galaxies}
\title{Looking for a link between Gas Content of Dwarf Galaxies and Properties of their Globular Cluster systems}
\author{Sharina, M.E.$^1$, Puzia, T.H.$^2$, Chandar R.$^3$, Goudfroij, P.$^4$, Davoust, E.$^5$}
\axindex{Sharina, M.E.}\axindex{Puzia, T.H.}\axindex{Chandar, R.}\axindex{Goudfroij, P.}\axindex{Davoust, E.}
\affil{ $^{1}$Special Astrophysical Observatory, Russian Academy of Sciences, N.Arkhyz, KChR, 369167, Russia\\
 $^{2}$Department of Physics and Astronomy, The University of Toledo, 2801 West
Bancroft Street, Toledo,
OH 43606\\
 $^{3}$Herzberg Institute of Astrophysics, 5071 West Saanich Road, Victoria, BC
V9E 2E7, Canada\\
 $^{4}$Space Telescope Science Institute, 3700 San Martin Drive, Baltimore, MD 21218\\
 $^{5}$Laboratoire d'Astrophysique de Toulouse-Tarbes, Universit\'e de Toulouse, CNRS, 14 avenue E.~Belin, F-31400, Toulouse, France}


\begin{abstract}
Star clusters are fundamental building blocks of galaxies. Their formation
is related to the density and pressure in progenitor molecular clouds and 
their environmental conditions. To understand better the dynamical 
processes driving star formation and chemical evolution, we compare ages,
metallicities, and alpha-element abundance ratios of
globular clusters in nearby dwarf galaxies of different luminosities
and morphological types, and situated in different environments. The data
are based on our 6m telescope medium-resolution spectroscopic observations.
We find that a mean metallicity of GCs in a galaxy at a given age is higher for
early-type dwarfs, than for late-type dwarf irregulars and spirals.
 \end{abstract}

\section{Introduction}
Low surface brightness (LSB) dwarf galaxies are fainter than $M_B=-16^m$ 
and they represent the most common type of objects in the local Universe.  
In distinction to globular clusters (GCs), they are dark-matter dominated.
Dwarf spheroidals (dSph) are gas-poor, old stellar systems. 
Dwarf irregulars (dIrr) contain young stars and neutral hydrogen, the fuel for
star formation activity. Some objects, classified as transitional types
(dIrr/dSph), show evidence for a few young and intermediate-age stars 
but not for any gas, or contain very small amounts of it, which is below
the detection level of most modern telescopes. 
The origin of dwarf galaxies is under debate. What processes 
drive morphological transformations? Why do some of these objects lose their gas? 
Clues to the solution of these problems can be found in the observational properties 
of these objects and their GC systems.

Ancient GCs are the brightest representatives of the oldest simple stellar populations.
Their younger cousins, massive compact star clusters, have formed in some galaxies over
the lifetime of the universe. 
The physical properties of both (hereafter: GCs) are often used for a better understanding of
the evolution of dwarf galaxies and their role as building blocks in the cosmological structure 
formation picture (see for a review West et al., 2004, and references therein).

\medskip

\section{Probability density estimates for evolutionary parameters of GCs in dwarf galaxies}
   In Fig.~1 we present probability density functions (PDF) for ages, metallicities, and [$\alpha$/Fe]
abundance ratios for GCs in dwarf galaxies obtained from our recent spectroscopic work. The evolutionary parameters
were derived from medium-resolution spectroscopic observations at the 6m telescope of 
the Russian Academy of Sciences (Sharina et al.~2010). We include in the analysis the data 
for GCs in dwarf galaxies obtained at the Very Large Telescope of the European Southern 
Observatory and the 6m telescope SAO RAS: Sharina, Afanasiev, and Puzia, 2006a,b; 
Sharina, Puzia \& Krylatyh, 2007; Puzia \& Sharina, 2008; Sharina \& Davoust, 2009.

For comparison, we use the data from the literature for GCs in more massive galaxies. 
Evolutionary parameters for 70 GCs in M31 were estimated by Puzia et al. (2005). 
[Fe/H] values were taken from the catalog of Harris (1996) for GCs in the Milky Way (MW). 
Ages and metallicities of 40 Galactic GCs were estimated by us using spectra from the sample of 
Schiavon et al. (2005). We obtain ages, metallicities, and [$\alpha$/Fe] ratios for 24 GCs 
in the Large Maggellanic Cloud (LMC) using the Lick index measurements of Beasley et al. (2002).  
Our measurements of evolutionary parameters for GCs in the LMC and the MW are summarized 
in Sharina et al. (2010). The lines in Fig.1 indicate non-parametric density estimates using an Epanechnikov kernel 
(Epanechnikov 1969). For an explanation of the diagram as a probability density distribution see Freedman \& Diaconis (1981).

Probability density estimates of [$\alpha$/Fe] for GCs in dwarf galaxies, the LMC and the MW are shown in the top 
right panel. The peaks are defined with the widths 0.08, 0.07, and 0.05 dex, respectively. It is seen, 
that the alpha-element abundance ratio is lower ([$\alpha/$Fe$]\!\sim\! 0.07$) in dwarf galaxies, 
than in the LMC ([$\alpha/$Fe$]\!\sim\!0.23$), and the MW ([$\alpha/$Fe$]\!\sim\! 0.4$ dex), although the dispersions are roughly 
of the same order.

PDFs of metallicity for the GCs are presented in the top, middle, and bottom left panels of Fig.1. 
The top panel demonstrates, that the median [Fe/H] value of our sample clusters for the LMC tends 
to be higher ([Fe/H]$\sim-0.8$ dex),
than for lower massive dwarfs ([Fe/H]$\sim-1.33$). The width of the non-parametric fit for the LMC is broader 
(0.38 dex). In the middle panel we consider independently GCs in LSB dIrrs, and old, and intermediate-age GCs in dSphs. It is seen that there are two narrow peaks ([Fe/H]$\sim-1.8$ and $-1.3$ dex) with widths roughly 0.05 dex for GCs in dIrrs, and two broad peaks ([Fe/H]$\sim-1.45$ and $-1.0$ dex) with widths $\sim\!0.2$ dex for dSphs. The values of metallicities and the width of the peaks are indications of the intensities of star formation in galaxies and their duration. The distribution of metallicities in the unified GC system of early-type dwarfs is as broad as the one of LMC, which is a low-mass spiral. 

The bottom panel represents the PDFs for GCs in the MW and M31, for comparison. The bimodality of the metallicity distributions, and the difference between the maximum [Fe/H] values between the two GC systems are clearly visible. The metallicity of the peak for dSphs is similar to the one for the MW. The M31 sample is mostly composed of GCs belonging to the metal-rich M31 bulge/disk sub-population (Puzia et al.~2005). It is seen, that this sub-system has the highest metallicity ([Fe/H]$\sim-0.6$) of its GC system among the samples considered here.  

The differences in the ages of the GC systems in dSphs, dIrrs, LMC, M31, and MW are seen in the middle, and bottom right panels of Fig.~1. 
The widths of the non-parametric density functions are a few Gyr in all the cases, and we note that the internal accuracy of the age determination is of the order $\Delta t/t\approx 0.3-0.5$. It is seen, that the GC system of our Galaxy is the oldest with a slightly younger sub-component (see also Mar{\'{\i}}n-Franch et al.~2009). Its age is roughly consistent with the one for dSphs. At this point, it is still unclear whether the age difference between the oldest MW bulge GC population and the oldest M31 is significant. The oldest GCs in dIrrs appear to be as old as intermediate-age GCs in LMC. Some LSB dwarfs contain clusters younger than 1 Gyr. 

\section{Conclusion}
We discuss the PDFs of evolutionary parameters (age, metallicity, [$\alpha$/Fe]) of GCs observed in our 6m telescope spectroscopic campaign and compare those with GCs from the literature. 
Our analysis shows that 
1) the metallicity and age dispersions in GC systems are wider for larger, i.e. more massive galaxies;
2) metal-rich clusters are preferentially found in galaxies more massive than $\sim\!10^9 M_{\sun}$;
3) intermediate-age GCs in early-type dwarf galaxies are richer in metals than star clusters representing dynamically
 "cold" gas-rich environments in dIrrs;

\begin{figure}[!h]
\centerline{\hbox{\psfig{figure=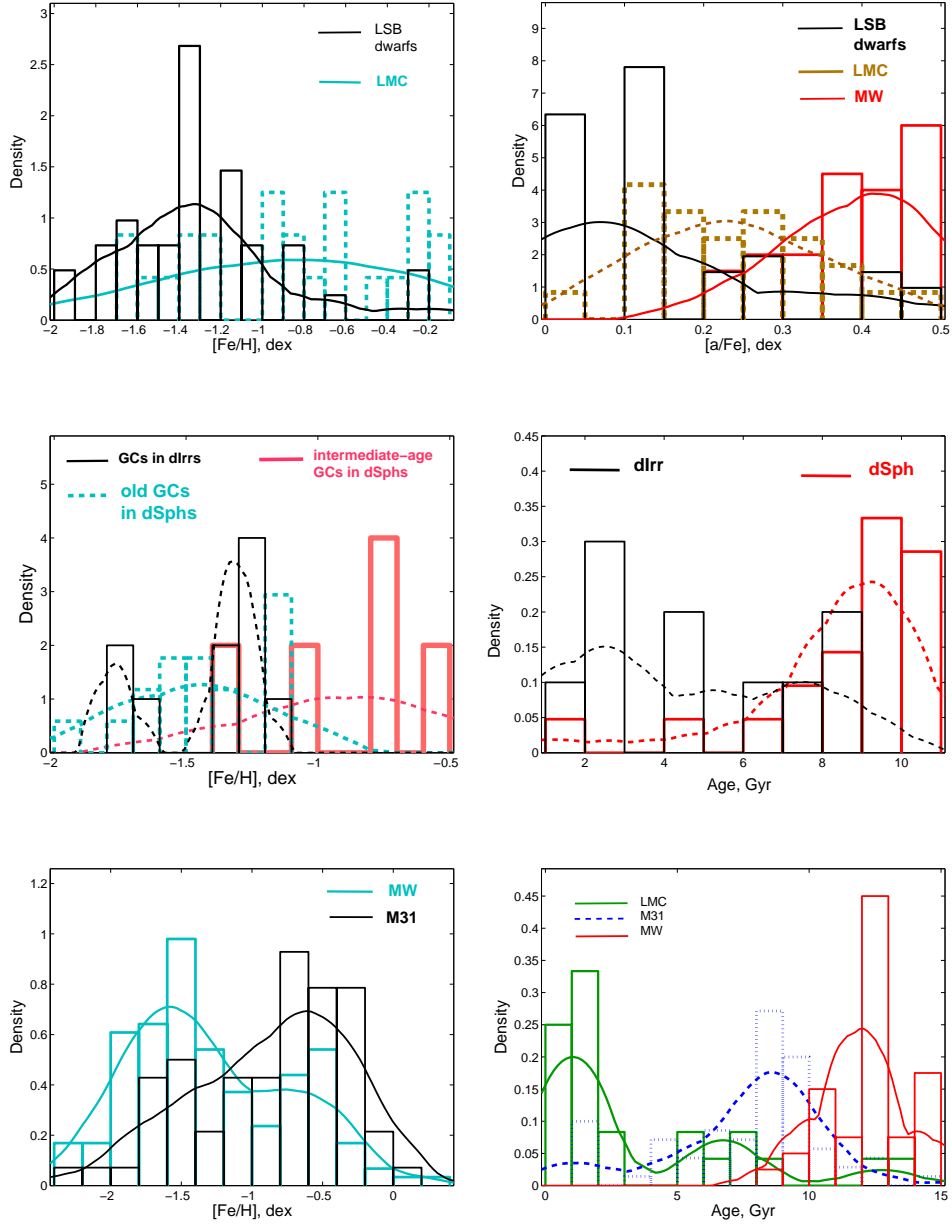,angle=0,clip=,width=13cm}}}
\caption[]{Probability density functions of evolutionary parameters (age, metallicity, [$\alpha$/Fe]) for GCs in
our Galaxy, M31, LMC, and dwarf galaxies in nearby groups. The curves indicate
non-parametric probability density estimates using an Epanechnikov kernel (see text for details).
}
\label{fig1}
\end{figure}



\end{document}